\documentclass[11pt]{article}
\usepackage[letterpaper]{geometry}

\usepackage{ltexpprt}
\usepackage[T1]{fontenc}
\usepackage{xcolor,amsmath,amsfonts,amssymb,graphicx}
\usepackage{algorithm}
\usepackage{algpseudocode}%
\usepackage[super]{nth}

\title{Interactive Verifiable Polynomial Evaluation}
\author{Salman Avestimehr}

\author{Saeid Sahraei\thanks{University of Southern California, Los Angeles, CA 90089, USA. Email: ss\_805@usc.edu}
\and Mohammad Ali Maddah-Ali\thanks{Nokia Bell Labs, Holmdel, NJ 07733, USA. Email: mohammad.maddahali@nokia-bell-labs.com}
\and Salman Avestimehr\thanks{University of Southern California, Los Angeles, CA 90089, USA. Email: avestimehr@ee.usc.edu}}

\date{}
\begin{document}
    \maketitle
\begin{abstract}
Cloud computing platforms have created the possibility for computationally limited users to delegate demanding tasks to strong but untrusted servers. Verifiable computing algorithms help build trust in such interactions by enabling the server to provide a proof of correctness of his results which the user can check very efficiently. In this paper, we present a doubly-efficient interactive algorithm for verifiable polynomial evaluation. Unlike the mainstream literature on verifiable computing, the soundness of our algorithm is information-theoretic and cannot be broken by a computationally unbounded server.  By relying on basic properties of error correcting codes, our algorithm enforces a dishonest server to provide false results to problems which become progressively easier to verify. After roughly $\log d$ rounds, the user can verify the response of the server against a look-up table that has been pre-computed during an initialization phase. For a polynomial of degree $d$, we achieve a user complexity of $O(d^{\epsilon})$, a server complexity of $O(d^{1+\epsilon})$, a round complexity of  $O(\log d)$ and an initialization complexity of $O(d^{1+\epsilon})$.
\end{abstract}

\section{Introduction}
The recent rise in cloud computing platforms has created an increasing demand for verifiable computing protocols. A commercial server that sells its computation power to the users, is incentivized to return false results, if by doing so it can reduce its computation overhead and oversell its services. Therefore, a user who delegates a computationally demanding task to a server must be able to verify that the results are indeed correct. Verifiable computing algorithms require the server to send a ``proof'' to the user along with the results of computation \cite{gennaro2010non}. By investigating this proof, a user is convinced with high probability that the server is honest. 
Clearly, a practical verifiable computing algorithm must be ``doubly-efficient'' \cite{goldwasser2015delegating}: it must have a super-efficient verifier (user) and an efficient prover (server). Put differently, if the complexity of computing the original function is $\chi$, the prover's complexity must be comparable to $\chi$, and the verifier's complexity must be substantially smaller than $\chi$. 

The study of verifiable computing has led to novel cryptographic algorithms which are either applicable to arbitrary functions \cite{gennaro2013quadratic,parno2013pinocchio} or tailored to computation of a specific function \cite{fiore2012publicly,benabbas2011verifiable}. The former has led to the development of Quadratic Arithmetic Programs and zkSNARKs for arithmetic circuits, while the latter has resulted in highly efficient and easy-to-implement algorithms that can be applied to, say, polynomial evaluation or matrix multiplication. 
Despite their enormous success, the security of cryptographic algorithms depends on hardness assumptions about certain mathematical problems. Algorithmic breakthroughs or technological advancements may falsify these assumptions at any time. This creates a demand for verifiable computing algorithms which remain secure in the face of computationally unbounded provers.

Our objective in this work is to design an information-theoretic algorithm for verifiable polynomial evaluation. Information-theoretic in this context means that the soundness of the algorithm is fundamental, and does not rely on hardness assumptions. An example of how such an algorithm can be applied in practice is blockchain networks where the capacity to verify a newly mined block against the history of the transactions is the distinguishing factor between a light node and a full node. This block verification process can generally be modelled as an instance of polynomial evaluation \cite{li2018polyshard}. The full nodes can thus convince the light nodes of the correctness of a block via a verifiable polynomial evaluation algorithm.

Our setup is very similar to the classical notion of interactive proof systems, with one subtle difference: we allow for a one-time initialization or pre-processing phase during which the verifier may perform a computationally heavy task. This is acceptable because in the verifiable computing literature, it is generally assumed \cite{gennaro2010non,benabbas2011verifiable,fiore2012publicly} that after this initialization phase, the verifier and the prover will engage in evaluating the function at many inputs. Therefore, this initialization cost amortizes over many rounds and can be neglected.
\subsection{Setting and Objective.}
\label{sec:objective}
Consider a polynomial of degree $d-1$, $f(x) = a_0 + a_1x+\cdots + a_{d-1}x^{d-1}$. A verifier wishes to evaluate this polynomial at $x\in\{x_1,x_2,\cdots\}$ with the help of a prover. Similar to \cite{goldwasser2015delegating}, we require the algorithm to be doubly-efficient with a small round complexity. On the other hand, we deviate from the classical notion of proof systems and allow the verifier to perform a one-time computationally heavy task. We impose the following performance criteria on the algorithm.
\begin{itemize}
\item {\bf Efficient initialization}: the verifier is allowed to perform a one-time {\it initialization} phase and store the outcome for future reference. Although this phase is run only once, its complexity should be comparable to the complexity of computing $f(x)$.
    \item {\bf Super-Efficient verifier}: for each $x\in\{x_1,x_2,\cdots\}$, the complexity of the verifier should be negligible compared to the complexity of computing $f(x)$.
    \item {\bf Efficient prover}: for each $x\in\{x_1,x_2,\cdots\}$, the complexity of the prover should be comparable to the complexity of computing $f(x)$.
    \item {\bf Small round complexity}: the number of rounds of interaction between the prover and the verifier must be polylogarithmic in $d$.
    \item {\bf Completeness}: if the prover is honest, the verifier should accept his results with probability $1$.
    \item {\bf (Information-theoretic) soundness}: If the prover is dishonest, the verifier should be able to reject his results with probability at least $\frac{1}{2}$.
\end{itemize}

\subsection{Related Work.}
An interactive proof (IP) system \cite{goldwasser1989knowledge,babai1985trading} is an interactive protocol that enables a prover to convince a computationally-limited verifier of the correctness of a statement. It is well-known that IPs are very strong tools: any problem in PSPACE admits an interactive proof with polynomial complexity for the verifier \cite{lund1990algebraic,shamir1990ip}. Nevertheless, the recent rise in cloud computing platforms has led to several intriguing questions surrounding the practicality of such algorithms. In particular, the provers in \cite{lund1990algebraic,shamir1990ip} run in exponential time, making them unfit for real-world commercial servers. The concept of doubly-efficient interactive proofs was first introduced in \cite{goldwasser2015delegating} followed by \cite{rothblum2013interactive,reingold2016constant}. 
In this context, ``doubly-efficient'' means that the prover must run in polynomial time and the verifier must be ``super-efficient'', i.e., his complexity must be close to linear in the size of the problem. Unfortunately, in many practical scenarios, even linear complexity is unacceptable for the verifier. Concrete examples of this are when the problem itself can be solved in linear time, however, due to the sheer size of the problem the verifier is incapable of performing the computation alone. In such cases, a high-degree polynomial-time prover is clearly impractical too. 

A closely related line of work is Probabilistically Checkable Proofs (PCP) \cite{ben2018fast,ben2016interactive,aszl1991checking,ben2008short,polishchuk1994nearly}, where the prover is required to commit to a proof which is usually too long for the verifier to process. The verifier can then sample this proof randomly in a few locations and be convinced of the correctness of the proof with high probability.  The celebrated PCP theorem \cite{arora1998proof,arora1998probabilistic} states that any problem in NP admits a PCP with verifier complexity that is polylogarithmic in the size of the problem.  However, from a practical perspective, it is not clear how this initial commitment can be implemented. One possibility is to rely on Merkle commitments with the help of collision-resistant hash functions and assume that the prover is computationally limited. Alternatively, the prover can send the entire proof to a trusted third-party which will be the point of contact for the verifier. However, these approaches alter the setup, and more importantly make strong assumptions such as the existence of trusted third parties or computational limitation of the prover. Other PCP-based proof systems that are secure against computationally limited provers include \cite{kilian1988founding} and \cite{micali1994cs}.

The notion of verifiable computing was introduced in \cite{gennaro2010non}. Motivated by practical considerations, in a verifiable computing setting one assumes that a computationally limited verifier (user) delegates a task to a prover (server). The prover must then cooperate with the verifier in computation of the function in such a way that the verifier remains efficient and convinced of the validity of the results. There are subtle differences between this model and the classical notion of proof systems. In particular, it is usually assumed that the verifier may run a one-time initialization phase that is computationally expensive. The cost of this computation is amortized over many runs of the algorithms, corresponding to the evaluation of the same function over many different inputs. In Gennaro's framework \cite{gennaro2010non}, the function to be computed is characterized by its boolean circuit which is evaluated with the help of Yao's garbled circuits \cite{yao1982protocols,lindell2009proof} and Fully Homomorphic Encryption. In a subsequent work \cite{gennaro2013quadratic,parno2013pinocchio} which led to the development of several zkSNARKs (zero-knowledge non-interactive arguments of knowledge) \cite{groth2016size,lipmaa2013succinct,bitansky2013succinct,sasson2014zerocash},  it was suggested to represent arbitrary arithmetic and logical circuits as Quadratic Arithmetic Programs and Quadratic Span Programs which are then evaluated at encrypted values to produce proofs of correctness. These algorithms can be proven secure against provers who are not powerful enough to reverse such encryptions. 

Recently, a new line of work in the cryptography community has focused on the verifiable evaluation of specific functions such as high-degree polynomials or multiplication of large matrices. These functions serve as building blocks for many applications such as machine learning and blockchain. Advances in this area has led to extremely efficient algorithms for verifiable polynomial evaluation \cite{benabbas2011verifiable, fiore2012publicly,backes2013verifiable,elkhiyaoui2016efficient}, matrix multiplication \cite{zhang2017new,fiore2012publicly}, and modular exponentiation \cite{chen2014new}. Unfortunately, similar to the generic verifiable computing algorithms discussed above, these algorithms rely on unproven cryptographic assumptions. To overcome this limitations, recent efforts have focused on information-theoretic verifiable polynomial evaluation algorithms which are secure against unbounded adversarial provers \cite{sahraei2019interpol}. Nevertheless, to achieve information-theoretic security, \cite{sahraei2019interpol} pays a significant cost in terms of the complexity of the verifier. While the cryptographic approaches lead to logarithmic or even constant verification time, the verifier in \cite{sahraei2019interpol} runs in $O(\sqrt{d})$.

\subsection{Main Results.}
we design an information-theoretic verifiable polynomial evaluation algorithm which is super-efficient for the verifier and efficient for the prover. Even though our algorithm is interactive, the number of interactions only grows logarithmically as a function of the degree of the polynomial. Similar to other verifiable computing algorithms, in our setting the verifier performs a one-time initialization phase whose complexity is comparable to the complexity of evaluating the polynomial once. 

\begin{theorem}
There exists a public-coin verifiable polynomial evaluation algorithm with initialization complexity of $O(d^{1+\epsilon})$, verifier complexity of $O(d^\epsilon)$, prover complexity of $O(d^{1+\epsilon})$, and round complexity of $O(\log d)$ which satisfies the completeness and soundness properties as defined in Section \ref{sec:objective}.
\end{theorem}

\subsection{An Overview of the Algorithm.}
Similar in nature to many other interactive proof systems \cite{lund1990algebraic,ben2018fast,ben2008short}, the core idea behind our algorithm is to enforce a dishonest prover to provide false results on problems that become progressively easier to verify throughout the interactions. To be more specific, let us look at a simple example where the polynomial $f(x) = a_0 + a_1x +\cdots + a_{d-1}x^{d-1}$ is to be evaluated where $d$ is even. Suppose the prover provides a false answer $\hat{f}(x) \neq f(x)$. The protocol then requires the prover to provide his evaluation of two more functions, namely $f^{(0)}(y) = a_0 + a_2 y + a_4y^2+\cdots + a_{d-2}y^{\frac{d}{2}-1}$ and $f^{(1)}(y) = a_1 + a_3 y + a_5y^2+\cdots + a_{d-1}y^{\frac{d}{2}-1}$, both evaluated at $y = x^2$. Obviously, we must have $f^{(0)}(y) + xf^{(1)}(y) = f(x)$, which can be easily checked by the verifier. Since $\hat{f}(x) \neq f(x)$, the prover must either provide $\hat{f}^{(0)}(y)\neq {f}^{(0)}(y)$ or  $\hat{f}^{(1)}(y)\neq {f}^{(1)}(y)$ in order to pass the verification $\hat{f}^{(0)}(y) + x\hat{f}^{(1)}(y) = \hat{f}(x)$. Voil\`{a}, the prover has been forced to lie again. Note also that the degree of $f^{(0)}$ and $f^{(1)}$ is only $d/2$. The verifier cloud then climb down the branches of this binary tree, and after $\log d$ rounds, verify the correctness of all the constant polynomials corresponding to the leaves of the tree. But this would take him  $O(d)$ to accomplish. Instead, he computes a random linear combination $\alpha \hat{f}^{(0)}(y)+\beta \hat{f}^{(1)}(y)$ and uses this as the reference point for the next iteration. Provided that these random coefficients are selected over sufficiently large sets, we will have $\alpha \hat{f}^{(0)}(y)+\beta \hat{f}^{(1)}(y)\neq \alpha {f}^{(0)}(y)+\beta {f}^{(1)}(y)$ with high probability and the error will propagate to the next iteration of the algorithm. After $\log d$ iterations, the verifier is left with a constant polynomial that he needs to verify. We note that this algorithm is inspired by the Probabilistically Checkable Proof of Proximity recently proposed in \cite{ben2018fast} (despite this, our algorithm is clearly not a PCP but an IP).

As it turns out, the fact that the degree of the polynomial decreases by a factor of $2$ at each iteration does not immediately translate to an easier verification. Unfortunately, each coefficient of the newly constructed polynomial is now a linear combination of two coefficients in the original polynomial. Therefore, evaluating the new polynomial from scratch still takes $O(d)$. In particular, after ${\log d}$ iterations, we are left with a constant which is a (random) linear combination of all the $d$ coefficients of the original polynomial. To help the verifier ensure the validity of this linear combination, we allow for an initialization phase, whereby the verifier computes all the possible outcomes of the algorithm and stores them in a look-up table in his memory. For this to be feasible, we need the random coefficients to be selected from a set that is not too large. By properly choosing the size of this set, we guarantee that the look-up table is of size $O(d^{1+\epsilon})$ and that it can be computed in time $O(d^{1+\epsilon})$. The verifier can then check the validity of the constant polynomial at the end of ${\log d}$ iterations by comparing the returned result to the corresponding entry in the look-up table.
\section{Interactive Verifiable Polynomial Evaluation}
In this section we describe an interactive algorithm for verifiable evaluation of a function $f(x) = a_0 + a_1x+\cdots+a_{d-1}x^{d-1}$ over members of a finite field $\mathbb{F}$. The coefficients $a_0,\cdots,a_{d-1}$ are arbitrary members of $\mathbb{F}$.
\subsection{Preliminaries.}
Let $\eta>1$ be an integer and let $c>1, c\in \mathbb{R}$ be such that $c\eta\in\mathbb{Z}$. These two are design parameters that can depend on $d$ in general. Throughout this section, we will assume that $c$ is a constant, and that $\eta$ grows at most polylogarithmically with $d$. Let $L\subseteq \mathbb{F}$ be a set of size $c \eta$ and let $H\subset L$ be of size $\eta$. Let us represent the members of $L$ as $L = \{\alpha_0,\cdots,\alpha_{c\eta-1}\}$. Without loss of generality, we can assume $H = \{\alpha_0,\cdots,\alpha_{\eta-1}\}$. The sets $L$ and $H$ are publicly known. For $i\in[0:\eta-1]$ define 
\begin{eqnarray}
Z_i(\beta) = \prod_{j\in[0:\eta-1]\backslash\{i\}}\frac{\beta - \alpha_j}{\alpha_i-\alpha_j}.
\end{eqnarray}
Note that for $i\in[0:\eta-1]$ and $k\in [0:\eta-1]$ we have $Z_i(\alpha_k) = 1$ if $i = k$ and $0$ if $i\neq k$. Let $r = \frac{\log d}{\log \eta}$. To simplify the matters, we assume $r$ is an integer (otherwise, we will set $r = \lceil\frac{\log d}{\log \eta}\rceil$). Iteratively, define the polynomials $g^{(b_1,\cdots,b_{\ell-1})}(\alpha,x)$ and $f^{(b_1,\cdots,b_\ell)}(x)$, for $\ell\in [0:r]$ and $(b_1,\cdots,b_{\ell})\in [0:c\eta-1]^{\ell}$ as follows. If $f^{(b_1,\cdots,b_{\ell-1})}(x) = e_0 + e_1 x + e_2 x^2 + \cdots$ then 
\begin{eqnarray}
g^{(b_1,\cdots,b_{\ell-1})}(\alpha,x) &=& Z_0(\alpha)[e_0 + e_{\eta}x + e_{2\eta}x^2 + \cdots]\\
&+& Z_1(\alpha)[e_1 + e_{\eta+1}x+ e_{2\eta+1}x^2+\cdots]\\
&+&\cdots\\
&+&Z_{\eta-1}(\alpha) [e_{\eta-1} + e_{2\eta-1}x+ e_{3\eta-1}x^2+\cdots],
\end{eqnarray}
and
\begin{eqnarray}
f^{(b_1,\cdots,b_{\ell})}(x) &=& g^{(b_1,\cdots,b_{\ell-1})}(\alpha_{b_\ell},x). 
\end{eqnarray}
As the starting point of the iteration, let  $f^{(b_1,\cdots,b_{\ell-1})}(x) =f(x)$ for $\ell= 1$. Note that $g^{(b_1,\cdots,b_{\ell})}(\alpha,x)$ is a polynomial of degree $\eta-1$ as a function of $\alpha$. Furthermore, note that 
\begin{eqnarray}
\sum_{b_\ell \in [0:\eta-1]}x^{b_\ell}f^{(b_1,\cdots,b_{\ell})}(x^\eta) = f^{(b_1,\cdots,b_{\ell-1})}(x).
\end{eqnarray}
Define $h(b_1,\cdots,b_{r})$ as
\begin{eqnarray}
h(b_1,\cdots,b_{r}) =  f^{(b_1,\cdots,b_{r})}(x), \;\; \forall (b_1,\cdots,b_\ell)\in[0:c\eta-1]^\ell.
\end{eqnarray}
Since $f^{(b_1,\cdots,b_{r})}(x)$ is a constant function of $x$, $h(b_1,\cdots,b_{r})$ only depends on $(b_1,\cdots,b_{r})$ and not $x$.

\subsection{The Initialization Phase.}
In the initialization phase, the verifier computes and stores $h(b_1,\cdots,b_{r})$ in a look-up table for all possible $(b_1,\cdots,b_r)\in[0:c\eta-1]^r$. In total, the number of evaluation points is 
\begin{eqnarray}
\lambda = (c\cdot \eta)^{r} = c^r\cdot d.
\label{eqn:lambda}
\end{eqnarray} 
We will now describe an algorithm for efficient computation of the look-up table. We will show that due to the recursive structure of $h(\cdot)$, its evaluation at all $\lambda$ points can be computed in $O(c^r \cdot d\cdot \eta)$.

\subsubsection{Efficient Computation of the Look-up Table.}
\label{sec:efficient_initialization}
A naive approach to obtaining the $\lambda = c^r \cdot d$ entries of the look-up table is to compute them individually, which would take $O(\lambda \cdot d)$. However, due to the recursive structure of the function $h(\cdot)$, we can be much more efficient. Specifically, let $a_i^{(b_1,\cdots,b_\ell)}$ be the coefficient of $x^i$ in $f^{(b_1,\cdots,b_\ell)}(x)$ for each $
\ell \in[0:r]$ and all $(b_1,\cdots,b_\ell)\in[0:c\eta-1]^{\ell}$. We have the following recursion 
\begin{eqnarray}
a_{i}^{(b_1,\cdots,b_\ell)} = \sum_{j\in[0:\eta-1]}a_{j+i\eta}^{(b_1,\cdots,b_{\ell-1})}Z_{j}(\alpha_{b_\ell}),\;\;\forall i\in[0:\frac{d}{\eta^\ell}-1]\;,\;\; (b_1,\cdots,b_\ell)\in[0:c\eta-1]^{\ell},
\label{eqn:iterative_a}
\end{eqnarray}
for all $\ell \in [1:r]$. Furthermore, for $\ell = 0$, we define $a_i^{(b_1,\cdots,b_\ell)} = a_i$. When $\ell = r$, the polynomial $f^{(b_1,\cdots,b_r)}(x)$ is of degree zero and is equal to $a_{0}^{(b_1,\cdots,b_r)}$. Therefore, we have
\begin{eqnarray}
h(b_1,\cdots,b_r) = a_{0}^{(b_1,\cdots,b_r)}.
\end{eqnarray}

 \begin{algorithm}
\caption[caption]{Efficient Computation of the Look-up Table}
\begin{algorithmic}[1]
\Statex {{\bf Input:} $(a_0,\cdots,a_{d-1})$, $\eta$, $c$, $L = \{\alpha_0,\cdots,\alpha_{c\eta-1}\}$.}
\Statex {\bf Output: }{The look-up table $h$}
\Statex{}
 \State $a^{()}_i = a_i$, for $i\in[0:d-1]$.
 \State $Z_{i,j} = \prod_{k\in[0:\eta-1]\backslash\{i\}}\frac{\alpha_j - \alpha_k}{\alpha_i-\alpha_k}$ for $i\in[0:\eta-1]$ and $j\in[0:c\eta-1]$.
\State $r = \lceil\frac{\log d }{\log \eta}\rceil$.
\For {$\ell\in[1:r]$}
\State $a_{i}^{(b_1,\cdots,b_\ell)} = \sum_{j\in[0:\eta-1]}a_{j+i\eta}^{(b_1,\cdots,b_{\ell-1})}Z_{j,b_\ell}\;\;$ for all $i\in[0:\frac{d}{\eta^\ell}-1]$ and $(b_1,\cdots,b_\ell)\in[0:c\eta-1]^{\ell}$.
\EndFor
\State $h(b_1,\cdots,b_r) = a_0^{(b_1,\cdots,b_r)}$ for all $(b_1,\cdots,b_r)\in[0:c\eta-1]^{r}$.
\State \Return $h$. 
\end{algorithmic}
\label{Alg:initialization}
\end{algorithm}

\begin{figure}
    \centering
    \includegraphics[width = 0.6\linewidth]{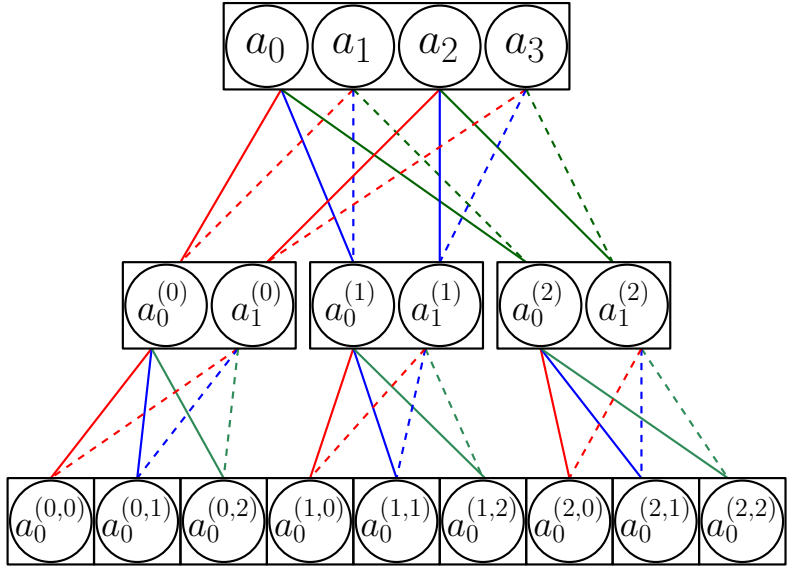}
    \caption{Efficient computation of the look-up table. Each box represents one polynomial, starting with $f(x) = a_0+a_1x+a_2x^2+a_3x^3$ at the root. The circles within each box represent the coefficients of the corresponding polynomial. Each edge is multiplied by $Z_i(\alpha_j)$ where $j$ determines the color (red = $0$, blue = $1$, green = $2$) and $i$ determines the style of the edge (solid for $i = 0$ and dashed for $i=1$). The two edges merging at each node are summed up to form the value of that node. At the base of the tree we have $h(b_1,b_2) = a^{(b_1,b_2)}_0$. In this example we have $d = 4, \eta = 2, c = \frac{3}{2}$.}
    \label{fig:setup_tree}
\end{figure}

In order to compute $h(b_1,\cdots,b_r)$, we start by computing a table of size $c\eta^2$ that stores all $Z_i(\alpha_j)$, for $i\in[0:\eta-1]$ and $j\in[0:c\eta-1]$ \footnote{We are assuming that $\eta$ grows at most polylogarithmically with $d$, so the complexity of computing $Z_i(\alpha_j)$ will be negligible compared to the overall complexity of the initialization phase. More on the proper choice of $\eta$ in Section \ref{sec:optimalchoice}.}. Afterwards, we iteratively compute the values of $a_i^{(b_1,\cdots,b_\ell)}$ until we reach the leaves of the tree which give us $h(b_1,\cdots,b_r)$. Since we have pre-computed the values of $Z_i(\alpha_j)$, finding each $a_i^{(b_1,\cdots,b_\ell)}$ based on \eqref{eqn:iterative_a} only takes $O(\eta)$. At the $\ell^{\textnormal{th}}$ level of the tree, we must compute $a_i^{(b_1,\cdots,b_\ell)}$ for all $(b_1,\cdots,b_\ell)\in[0:c\eta-1]^\ell$ and all $i\in[0:\frac{d}{\eta^\ell}-1]$. Therefore, the computation at level $\ell$ takes $O((c\cdot\eta)^{\ell}\cdot\frac{d}{\eta^\ell}\cdot\eta) = O({c^\ell\cdot d}\cdot\eta)$ operations. As a result, the entire tree can be computed in $c\cdot d\cdot\eta + c^2\cdot d\cdot\eta  +\cdots + c^r\cdot d \cdot \eta \le c^r\cdot d\cdot\eta\cdot\frac{c}{c-1} = O(c^r\cdot d\cdot\eta)$. This procedure has been summarized in Algorithm \ref{Alg:initialization}. For an illustration, see Figure \ref{fig:setup_tree}.

\subsection{The Evaluation Phase.}
\label{sec:evaluation}
\begin{figure}
    \centering
    \includegraphics[width=\linewidth]{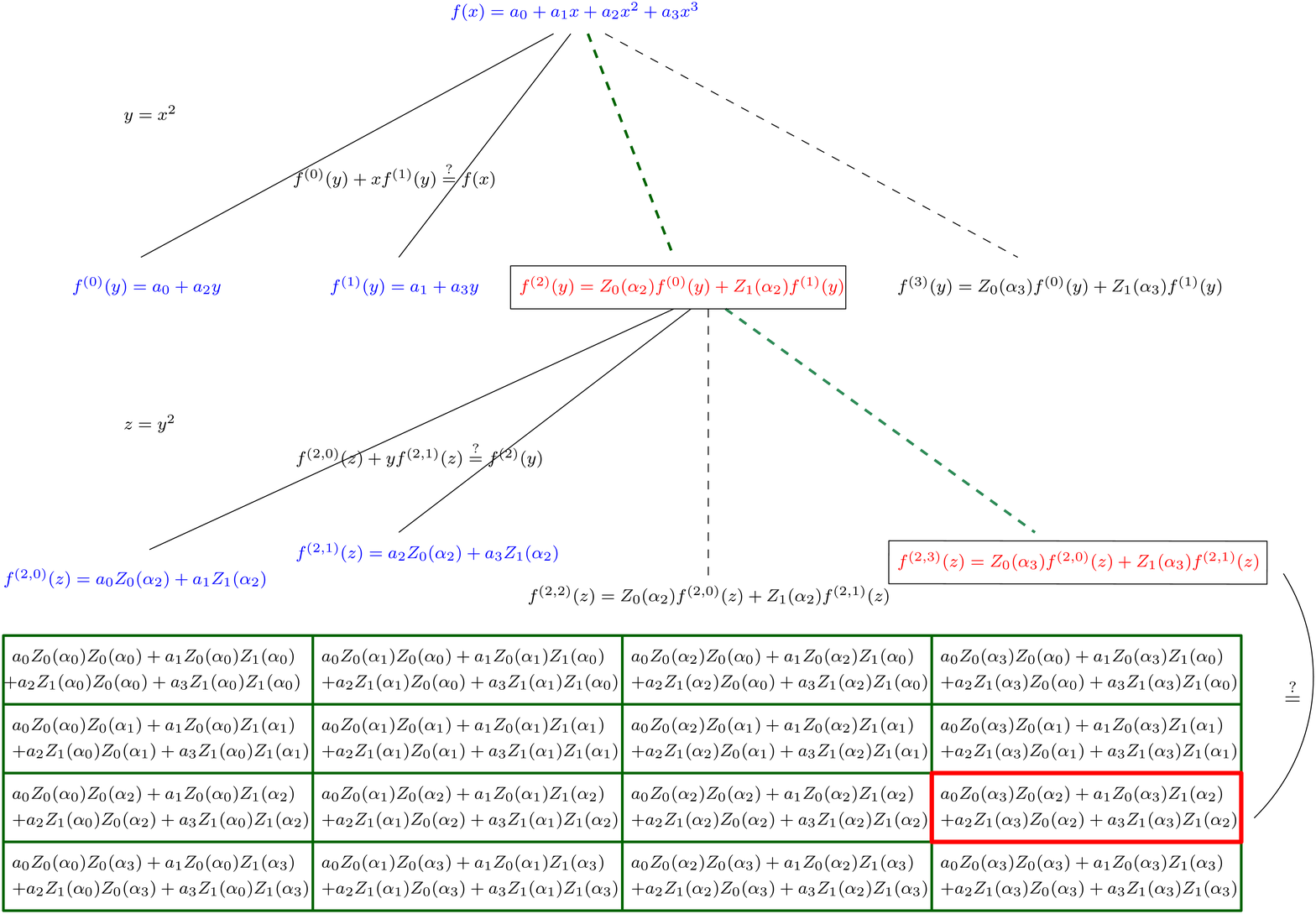}
    \caption{An illustration of the proposed algorithm. The blue values are provided by the prover. The red boxed values are computed by the verifier after interpolating the blue values at the same level. The verifier checks whether consecutive levels of the tree are consistent (identities marked by $\stackrel{?}{=}$). At the lowest level, the boxed red value is compared against a pre-computed look-up table. In this example we have $d = 4, \eta = 2, c =2$.}
    \label{fig:illustration}
\end{figure}

The verifier is interested in evaluating $f(x) = a_0 + a_1x+\cdots + a_{d-1}x^{d-1}$. We assume that both $f(\cdot)$ and $x$ are publicly known.
The prover sends $\hat{f}(x)$ to the verifier. If the prover is honest, then $\hat{f}(x) = f(x)$. Otherwise, it can be an arbitrary member of $\mathbb{F}$. The prover also sends the verifier $\hat{f}^{{(s)}}(x^{(1)})$ for all $s\in [0:\eta -1]$, where $x^{(1)} = x^\eta$. The verifier checks whether 
\begin{eqnarray}
\sum_{s\in [0:\eta-1]}x^{s}\hat{f}^{(s)}(x^{(1)}) = \hat{f}(x).
\label{eqn:firstcheck}
\end{eqnarray}
If not, he rejects the result. Next, the verifier finds the polynomial $\hat{g}(\alpha,x^{(1)})$ of degree $\eta-1$ (in $\alpha$) by interpolating the points $(\alpha_{s},\hat{f}^{{(s)}}(x^{(1)}))$, $s\in[0:\eta-1]$.  He then chooses $b_1\in[0:c\eta-1]$ uniformly at random and finds $\hat{f}^{(b_1)}(x^{(1)}) = \hat{g}(\alpha_{b_\ell},x^{(1)})$. The verifier then sends $b_1$ to the prover.  Next, the prover sends the verifier $\hat{f}^{(b_1,s)}(x^{(2)})$ for all $s \in[0:\eta-1]$ where $x^{(2)} = x^{\eta^2}$. The verifier checks if
\begin{eqnarray}
\sum_{s\in [0:\eta-1]}x^{\eta s}\hat{f}^{(b_1,s)}(x^{(2)}) = \hat{f}^{(b_1)}(x^{(1)}).
\end{eqnarray}
If not, he rejects the result. The algorithm now proceeds with $\hat{f}^{(b_1)}(x^{(1)})$ taking the role of $\hat{f}(x)$. This process continues until the prover sends the verifier $\hat{h}(b_1,\cdots,b_{r}) =  \hat{f}^{(b_1,\cdots,b_{r})}(x^{(r)}).$ The verifier checks this against the correct value of ${h}(b_1,\cdots,b_{r})$ stored in his look-up table. If they are not equal, the result is rejected. To amplify the prover's probability of failure, this entire algorithm is run $m$ times. If all the $m$ experiments pass, the verifier accepts the result. We will see in Section \ref{sec:performance} that the proper choice of $m$ is $(\frac{c}{c-1})^r$. This process has been summarized in Algorithm \ref{Alg:computation} and illustrated in Figure \ref{fig:illustration}. For notational simplicity, the algorithm is described as $m$ consecutive rounds. But the $m$ rounds can be trivially parallelized to reduce the overall round complexity to $r$.

 \begin{algorithm}
\caption[caption]{Interactive Verifiable Polynomial Evaluation, Verifier's Protocol}
\begin{algorithmic}[1]
\Statex {{\bf Input:} $x$, $\eta$, $c$, $d$, $Z_{i,j}$ for $i\in[0:\eta-1], j\in[0:c\eta-1]$ and the look-up table $h$.}
\Statex {\bf Output: }{One bit, accept or reject}
\Statex{}
\State $r = \lceil\frac{\log d }{\log \eta}\rceil$.
 \State $\hat{f}^{()} =$Input$()$.
%
\For {$i\in [1:m]$}
\For {$\ell\in[1:r]$}
\State $\hat{f}^{(b_1,\cdots,b_{\ell-1},s)} =$Input() for all $s\in[0:\eta-1]$.
 \If{$\hat{f}^{(b_1,\cdots,b_{\ell-1})}\neq \sum_{s\in [0:\eta-1]}x^{s}\hat{f}^{(b_1,\cdots,b_{\ell-1},s)} $}
 \State \Return reject.
 \EndIf
  \State Choose $b_\ell\sim$Uniform($[0:c\eta-1]$) and reveal $b_\ell$ to the prover.
\State Compute $\hat{f}^{(b_1,\cdots,b_\ell)} = \sum_{s\in[0:\eta-1]}Z_{s,b_\ell}\hat{f}^{(b_1,\cdots,b_{\ell-1},s)}$.
 \State Set $x = x^\eta$.
\EndFor
\If{$\hat{f}^{(b_1,\cdots,b_r)}\neq h(b_1,\cdots,b_r) $} \State \Return reject.
\EndIf
\EndFor
\State \Return accept.
\end{algorithmic}
\label{Alg:computation}
\end{algorithm}

\section{Performance Analysis}
\label{sec:performance}
\noindent{\bf Completeness}: If the prover is honest, he will provide the correct $f(x)$ and the correct values of $f^{(b_1,\cdots,b_\ell)}(x^{(\ell)})$ for all $\ell\in[1:r]$ which will clearly pass all the verification tests.

\noindent{\bf (Information-theoretic) soundness}:
Soundness follows from the simple principle that two distinct polynomials of degree $\eta-1$ must disagree on at  least $(c-1)\eta + 1$ points on any set of size $c\eta$. Suppose the prover starts by sending the verifier the wrong value of $\hat{f}(x)$. In order to pass \eqref{eqn:firstcheck}, he must then provide the verifier with at least one wrong value $\hat{f}^{(s)}(x_1)$. Because of this, the polynomial $\hat{g}(\alpha,x_1)$, as a function of $\alpha$, will be distinct from the correct polynomial ${g}(\alpha,x_1)$. Due to the observation above, these two polynomials will differ on at least $(c-1)\eta+1$ members of $L$. Therefore, if $b_1$ is chosen uniformly at random over $[0:c\eta-1]$, the value of $\hat{f}^{(b_1)}(x_1) = \hat{g}(\alpha_{b_1},x_1)$ will be different from the correct value ${f}^{(b_1)}(x_1)$ with probability at least $\frac{c-1}{c}$. With high probability, the error continues to propagate through the interactions between the verifier and the prover, until it reaches level $r$, at which point, the verifier can detect it by checking it against its stored value ${h}(b_1,\cdots,b_{r})$. The probability that an adversarial prover can successfully pass all the verifications is bounded by 
\begin{eqnarray}
p \le \frac{1}{c} + \frac{c-1}{c}\cdot\frac{1}{c} + (\frac{c-1}{c})^2\cdot\frac{1}{c} + \cdots + \cdot(\frac{c-1}{c})^{r-1}\cdot\frac{1}{c} = 1 - (1-\frac{1}{c})^{r}.
\end{eqnarray}
The verifier proceeds to run the experiment $m$ times and rejects the result if any of the $m$ experiments fail. We want to choose $m$ such that $p^m < 1/2$. By choosing $m = (\frac{c}{c-1})^{r}$, we will have
\begin{eqnarray}
p^m = (1-\frac{1}{m})^m < \frac{1}{e} < \frac{1}{2}.
\end{eqnarray}

\noindent{\bf Computational Complexity of the Initialization}: Based on the analysis in Section \ref{sec:efficient_initialization}, the initialization phase can be done in time 
\begin{eqnarray}
{\cal C}_{ini} =  O(c^\frac{\log d}{\log\eta}\cdot d\cdot \eta).
\end{eqnarray}
\noindent{\bf Computational Complexity of the Verifier}:
The verifier runs $m$ (parallel) experiments, each consisting of $r$ rounds. Each round takes $O(\eta)$ operations. Therefore, the overall complexity of the verifier is 
\begin{eqnarray}
{\cal C}_{ver} = O(m\cdot r\cdot \eta ) = O(\eta\cdot(\frac{c}{c-1})^{\frac{\log d}{\log \eta}}\cdot \frac{\log d}{\log\eta}).
\end{eqnarray}
\noindent{\bf Computational Complexity of the Prover}: An honest prover can also pre-compute the coefficients of the polynomials $f^{(b_1,\cdots,b_\ell)}(x)$ in its own initialization phase and store them locally in order to reduce its complexity. In round $\ell$, the prover must evaluate $\eta$ polynomials each of degree $\frac{d}{\eta^\ell}$. Therefore, the prover only needs to perform $m(d+\frac{d}{\eta}\cdot \eta + \frac{d}{\eta^2}\cdot \eta + \cdots + \frac{d}{\eta^r}\cdot\eta) \le m\cdot d\cdot\frac{2\eta}{\eta-1} = O(m\cdot d)$ computation \footnote{Even in the absence of an initialization phase for the prover, the complexity remains $O(m\cdot d\cdot \eta)$ for each of the $r$ rounds, resulting in an overall complexity of $O(m\cdot d\cdot \eta\cdot r)$. This is still of the form $O(d^{1+\epsilon})$ for the choice of $\eta$ and $c$ in Section \ref{sec:optimalchoice}.}.
\begin{eqnarray}
{\cal C}_{pro} = O(m\cdot d) = O(d\cdot (\frac{c}{c-1})^{\frac{\log d}{\log \eta}}).
\end{eqnarray}

\noindent{\bf Round Complexity of the Algorithm}: As mentioned in Section \ref{sec:evaluation}, we can run all the $m$ experiments in parallel. As a result, the round complexity of the algorithm is only 

\begin{eqnarray}
{\cal C}_{rnd} = \frac{\log d}{\log \eta}.
\end{eqnarray}


\subsection{The Proper Choice of the Parameters.}
\label{sec:optimalchoice}
We have two parameters to play with, namely $c$ and $\eta$. Firstly, we want to make sure that the initialization phase can be done in ${O}(d^{1+\epsilon})$. For this purpose, we choose $\eta = (\log d)^\omega$ for an arbitrary real number $\omega > 0$. We also fix $c\in\mathbb{R}$, $c>1$ to be a constant. To see why ${\cal C}_{ini} = {O}(d^{1+\epsilon})$, note that
\begin{eqnarray*}
{\cal C}_{ini} = c^{r}\cdot\eta \cdot d &=& c^{\frac{\log d}{\log \eta}}\cdot \eta\cdot d = d^{\frac{\log c}{\omega\log
\log d}}\cdot d\cdot (\log d)^\omega = {O}(d^{1+\frac{\log c}{\omega\log\log d}}).
\end{eqnarray*}
Our second criterion is to achieve ${\cal C}_{ver} = O(d^\epsilon)$. This requirement is automatically satisfied with the above choices of $\eta$ and $c$.
\begin{eqnarray}
{\cal C}_{ver} = O(\eta\cdot (\frac{c}{c-1})^\frac{\log d}{\log \eta}\cdot \frac{\log d}{\log \eta}) =O( d^{\frac{b}{\log\log d}}\cdot \frac{(\log d)^{\omega+1}}{\omega \log\log d}) = O(d^{\frac{b}{\log\log d}}),
\end{eqnarray}
where $b = \frac{1}{\omega}\cdot\log \frac{c}{c-1}$ is a constant. Finally, the complexity of the prover is given by 
\begin{eqnarray}
{\cal C}_{pro} = O(d\cdot(\frac{c}{c-1})^\frac{\log d}{\log \eta}) = O(d^{1+\frac{b}{\log\log d}}),
\end{eqnarray}
and the round complexity is 
\begin{eqnarray}
{\cal C}_{rnd} = \frac{\log d}{\log \eta} = O(\log d).
\end{eqnarray}

\section{Discussion: Extending the Results to Multivariate Polynomials}
Consider an $n$-variate polynomial of degree $d-1$ in each variable
\begin{eqnarray}
f(x_1,\cdots,x_n) = \sum_{i_1\in[0:d-1]}\cdots\sum_{i_n\in[0:d-1]}a_{i_1,\cdots,i_n}\prod_{j\in[1:n]}x_j^{i_j}.
\end{eqnarray}
A verifier wishes to evaluate this polynomial at $(x_1,\cdots,x_n)$ with the help of a prover. A simple variation of Algorithm \ref{Alg:computation} can be used for this purpose. First, the verifier treats $f(x_1,\cdots,x_n)$ as a univariate polynomial in $x_1$ and applies Algorithm \ref{Alg:computation} in order to reduce the degree of $x_1$ to zero after $r = \frac{\log d}{\log \eta}$ interactions with the prover. Now, he is left with a new polynomial that only has $n-1$ variables. After $n\cdot r$ rounds, the number of variables will reduce to zero, and the verifier will be left with a constant that he can check against a look-up table of size $(c\cdot \eta)^{nr}$. It is easy to see that if we resort to this modified algorithm, all the results in Section \ref{sec:performance} will remain valid, except $d$ must be replaced with $d^n$ (the number of terms in $f(x_1,\cdots,x_n)$). For instance, the complexity of the verifier will be
\begin{eqnarray}
{\cal C}_{ver,n} = O(d^{\frac{bn}{\log n + \log\log d}})
\end{eqnarray}
and the complexity of the prover will be 
\begin{eqnarray}
{\cal C}_{pro,n} = O(d^{1 + \frac{bn}{\log n + \log\log d}})
\end{eqnarray}
It is also noteworthy that any improvements in the multivariate case directly translates to a better univariate algorithm. To see why, consider an arbitrary univariate polynomial $g(x)$ of degree $d^n-1$. Without loss of generality, we can represent this polynomial as 
\begin{eqnarray}
g(x) = \sum_{i_1\in[0:d-1]}\cdots\sum_{i_n\in[0:d-1]}a_{i_1,\cdots,i_n} x^{i_1 + i_2d +\cdots + i_nd^{n-1}} = f(x,x^d,\cdots,x^{d^{n-1}}).
\end{eqnarray}
Applying Algorithm \ref{Alg:computation} on $g(x)$ results in a verifier complexity of $O(d^{\frac{bn}{\log n + \log\log d}})$ which is the same as ${\cal C}_{ver,n}$. If we can design an $n-$variate algorithm that achieves a verifier complexity smaller than ${\cal C}_{ver,n}$, we can apply it on $f(x,x^d,\cdots,x^{d^{n-1}})$ and improve upon the complexity of Algorithm \ref{Alg:computation} applied on $g(x)$.
\section*{Acknowledgement}
The authors would like to thank Ali Rahimi for the fruitful discussions. 
\bibliographystyle{acm}
\bibliography{main.bib}

\begin{thebibliography}{10}

\bibitem{arora1998proof}
{\sc Arora, S., Lund, C., Motwani, R., Sudan, M., and Szegedy, M.}
\newblock Proof verification and the hardness of approximation problems.
\newblock {\em Journal of the ACM (JACM) 45}, 3 (1998), 501--555.

\bibitem{arora1998probabilistic}
{\sc Arora, S., and Safra, S.}
\newblock Probabilistic checking of proofs: A new characterization of np.
\newblock {\em Journal of the ACM (JACM) 45}, 1 (1998), 70--122.

\bibitem{aszl1991checking}
{\sc aszl~o Babai, L., Fortnow, L., Levin, L., and Szegedy, M.}
\newblock Checking computations in polylogarithmic time.
\newblock In {\em 23rd Annual ACM Symposium on Theory of Computing\/} (1991),
  ACM, pp.~21--31.

\bibitem{babai1985trading}
{\sc Babai, L.}
\newblock Trading group theory for randomness.
\newblock In {\em Proceedings of the seventeenth annual ACM symposium on Theory
  of computing\/} (1985), ACM, pp.~421--429.

\bibitem{backes2013verifiable}
{\sc Backes, M., Fiore, D., and Reischuk, R.~M.}
\newblock Verifiable delegation of computation on outsourced data.
\newblock In {\em Proceedings of the 2013 ACM SIGSAC conference on Computer \&
  communications security\/} (2013), ACM, pp.~863--874.

\bibitem{ben2018fast}
{\sc Ben-Sasson, E., Bentov, I., Horesh, Y., and Riabzev, M.}
\newblock Fast {R}eed-{S}olomon interactive oracle proofs of proximity.
\newblock In {\em 45th International Colloquium on Automata, Languages, and
  Programming (ICALP 2018)\/} (2018), Schloss Dagstuhl-Leibniz-Zentrum fuer
  Informatik.

\bibitem{ben2016interactive}
{\sc Ben-Sasson, E., Chiesa, A., and Spooner, N.}
\newblock Interactive oracle proofs.
\newblock In {\em Theory of Cryptography Conference\/} (2016), Springer,
  pp.~31--60.

\bibitem{ben2008short}
{\sc Ben-Sasson, E., and Sudan, M.}
\newblock Short {PCP}s with polylog query complexity.
\newblock {\em SIAM Journal on Computing 38}, 2 (2008), 551--607.

\bibitem{benabbas2011verifiable}
{\sc Benabbas, S., Gennaro, R., and Vahlis, Y.}
\newblock Verifiable delegation of computation over large datasets.
\newblock In {\em Annual Cryptology Conference\/} (2011), Springer,
  pp.~111--131.

\bibitem{bitansky2013succinct}
{\sc Bitansky, N., Chiesa, A., Ishai, Y., Paneth, O., and Ostrovsky, R.}
\newblock Succinct non-interactive arguments via linear interactive proofs.
\newblock In {\em Theory of Cryptography Conference\/} (2013), Springer,
  pp.~315--333.

\bibitem{chen2014new}
{\sc Chen, X., Li, J., Ma, J., Tang, Q., and Lou, W.}
\newblock New algorithms for secure outsourcing of modular exponentiations.
\newblock {\em IEEE Transactions on Parallel and Distributed Systems 25}, 9
  (2014), 2386--2396.

\bibitem{elkhiyaoui2016efficient}
{\sc Elkhiyaoui, K., {\"O}nen, M., Azraoui, M., and Molva, R.}
\newblock Efficient techniques for publicly verifiable delegation of
  computation.
\newblock In {\em Proceedings of the 11th ACM on Asia Conference on Computer
  and Communications Security\/} (2016), ACM, pp.~119--128.

\bibitem{fiore2012publicly}
{\sc Fiore, D., and Gennaro, R.}
\newblock Publicly verifiable delegation of large polynomials and matrix
  computations, with applications.
\newblock In {\em Proceedings of the 2012 ACM conference on Computer and
  communications security\/} (2012), ACM, pp.~501--512.

\bibitem{gennaro2010non}
{\sc Gennaro, R., Gentry, C., and Parno, B.}
\newblock Non-interactive verifiable computing: Outsourcing computation to
  untrusted workers.
\newblock In {\em Annual Cryptology Conference\/} (2010), Springer,
  pp.~465--482.

\bibitem{gennaro2013quadratic}
{\sc Gennaro, R., Gentry, C., Parno, B., and Raykova, M.}
\newblock Quadratic span programs and succinct {NIZK}s without {PCP}s.
\newblock In {\em Annual International Conference on the Theory and
  Applications of Cryptographic Techniques\/} (2013), Springer, pp.~626--645.

\bibitem{goldwasser2015delegating}
{\sc Goldwasser, S., Kalai, Y.~T., and Rothblum, G.~N.}
\newblock Delegating computation: interactive proofs for muggles.
\newblock {\em Journal of the ACM (JACM) 62}, 4 (2015), 27.

\bibitem{goldwasser1989knowledge}
{\sc Goldwasser, S., Micali, S., and Rackoff, C.}
\newblock The knowledge complexity of interactive proof systems.
\newblock {\em SIAM Journal on computing 18}, 1 (1989), 186--208.

\bibitem{groth2016size}
{\sc Groth, J.}
\newblock On the size of pairing-based non-interactive arguments.
\newblock In {\em Annual International Conference on the Theory and
  Applications of Cryptographic Techniques\/} (2016), Springer, pp.~305--326.

\bibitem{kilian1988founding}
{\sc Kilian, J.}
\newblock Founding cryptography on oblivious transfer.
\newblock In {\em Proceedings of the twentieth annual ACM symposium on Theory
  of computing\/} (1988), ACM, pp.~20--31.

\bibitem{li2018polyshard}
{\sc Li, S., Yu, M., Avestimehr, S., Kannan, S., and Viswanath, P.}
\newblock {PolyShard}: Coded sharding achieves linearly scaling efficiency and
  security simultaneously.
\newblock {\em arXiv preprint arXiv:1809.10361\/} (2018).

\bibitem{lindell2009proof}
{\sc Lindell, Y., and Pinkas, B.}
\newblock A proof of security of {Y}ao’s protocol for two-party computation.
\newblock {\em Journal of Cryptology 22}, 2 (2009), 161--188.

\bibitem{lipmaa2013succinct}
{\sc Lipmaa, H.}
\newblock Succinct non-interactive zero knowledge arguments from span programs
  and linear error-correcting codes.
\newblock In {\em International Conference on the Theory and Application of
  Cryptology and Information Security\/} (2013), Springer, pp.~41--60.

\bibitem{lund1990algebraic}
{\sc Lund, C., Fortnow, L., Karloff, H., and Nisan, N.}
\newblock Algebraic methods for interactive proof systems.
\newblock In {\em Proceedings [1990] 31st Annual Symposium on Foundations of
  Computer Science\/} (1990), IEEE, pp.~2--10.

\bibitem{micali1994cs}
{\sc Micali, S.}
\newblock {C}omputationally {S}ound proofs.
\newblock In {\em Proceedings 35th Annual Symposium on Foundations of Computer
  Science\/} (1994), IEEE, pp.~436--453.

\bibitem{parno2013pinocchio}
{\sc Parno, B., Howell, J., Gentry, C., and Raykova, M.}
\newblock Pinocchio: Nearly practical verifiable computation.
\newblock In {\em 2013 IEEE Symposium on Security and Privacy\/} (2013), IEEE,
  pp.~238--252.

\bibitem{polishchuk1994nearly}
{\sc Polishchuk, A., and Spielman, D.~A.}
\newblock Nearly-linear size holographic proofs.
\newblock In {\em Proceedings of the twenty-sixth annual ACM symposium on
  Theory of computing\/} (1994), ACM, pp.~194--203.

\bibitem{reingold2016constant}
{\sc Reingold, O., Rothblum, G.~N., and Rothblum, R.~D.}
\newblock Constant-round interactive proofs for delegating computation.
\newblock In {\em Proceedings of the forty-eighth annual ACM symposium on
  Theory of Computing\/} (2016), ACM, pp.~49--62.

\bibitem{rothblum2013interactive}
{\sc Rothblum, G.~N., Vadhan, S., and Wigderson, A.}
\newblock Interactive proofs of proximity: delegating computation in sublinear
  time.
\newblock In {\em Proceedings of the forty-fifth annual ACM symposium on Theory
  of computing\/} (2013), ACM, pp.~793--802.

\bibitem{sahraei2019interpol}
{\sc Sahraei, S., and Avestimehr, A.~S.}
\newblock {INTERPOL}: Information theoretically verifiable polynomial
  evaluation.
\newblock {\em arXiv preprint arXiv:1901.03379\/} (2019).

\bibitem{sasson2014zerocash}
{\sc Sasson, E.~B., Chiesa, A., Garman, C., Green, M., Miers, I., Tromer, E.,
  and Virza, M.}
\newblock Zerocash: Decentralized anonymous payments from bitcoin.
\newblock In {\em 2014 IEEE Symposium on Security and Privacy\/} (2014), IEEE,
  pp.~459--474.

\bibitem{shamir1990ip}
{\sc Shamir, A.}
\newblock {IP= PSPACE} (interactive proof= polynomial space).
\newblock In {\em Proceedings [1990] 31st Annual Symposium on Foundations of
  Computer Science\/} (1990), IEEE, pp.~11--15.

\bibitem{yao1982protocols}
{\sc Yao, A.~C.}
\newblock Protocols for secure computations.
\newblock In {\em Foundations of Computer Science, 1982. SFCS'08. 23rd Annual
  Symposium on\/} (1982), IEEE, pp.~160--164.

\bibitem{zhang2017new}
{\sc Zhang, X., Jiang, T., Li, K.-C., Castiglione, A., and Chen, X.}
\newblock New publicly verifiable computation for batch matrix multiplication.
\newblock {\em Information Sciences\/} (2017).

\end{thebibliography}
\end{document}